\documentclass[fleqn,usenatbib]{mnras}
\usepackage{ulem}
\usepackage[utf8]{inputenc}
\usepackage{xcolor}

\newcommand{\II}{\small{II}\normalsize}

\usepackage{newtxtext,newtxmath}

\usepackage[T1]{fontenc}
\usepackage{ae,aecompl}
\usepackage{siunitx}
\usepackage{soul}
\usepackage{longtable}
\providecommand{\abs}[1]{\lvert#1\rvert}

\usepackage{graphicx}	
\usepackage{amsmath}	
\usepackage{enumerate}
\usepackage{natbib}
\usepackage{adjustbox}
\usepackage{helvet,soul}
\usepackage[font=small]{caption}

\title[Activity Cycles in RS CVn]{Activity cycles in RS-CVn Stars}

\author[C. I. Mart\'inez]{
C. I. Mart\'inez$^{1,2}$\thanks{Contact e-mail: \href{mailto:cmartinez@conicet.gov.ar}{cmartinez@conicet.gov.ar}}, P. J. D. Mauas$^{3,4}$\thanks{Contact e-mail: \href{mailto:pablo@iafe.uba.ar}{pablo@iafe.uba.ar}}, A. P. Buccino$^{3,4}$ \\
\\
$^{1}$Observatorio Astron\'omico F\'elix Aguilar, Universidad Nacional de San Juan, Av. Benavides 8175 oeste, CP 5400 San Juan, Argentina.\\
$^{2}$Instituto de Ciencias Astron\'omicas de la Tierra y el Espacio (CONICET-UNSJ), Av. España 1512 sur, San Juan, Argentina.\\
$^{3}$Instituto de Astronom\'\i a y F\'\i sica del Espacio (CONICET-UBA), C.C. 67 Sucursal 28, C1428EHA-Buenos Aires, Argentina. \\
$^{4}$Departamento de F\'\i sica. Facultad de Ciencias Exactas y Naturales - Universidad de Buenos Aires, Buenos, Argentina.\\
}

\pubyear{2021}

\begin{document}
\label{firstpage}
\pagerange{\pageref{firstpage}--\pageref{lastpage}}
\maketitle

\begin{abstract}

We compile a list of 121 RS CVn type stars from the bibliography in southern hemisphere, to search for activity cycles, covering a large range of luminosities and rotation periods. For each system of the list, we obtain photometric data from the All Sky Automated Survey (ASAS), and we complement it with our own photometry, obtained with the Optical Robotic Observatory (ORO). We analyze this data with the Generalized Lomb–Scargle periodogram to determine the rotation period and possible activity cycles for each system. We found rotation periods for 102 systems and activity cycles for 91 systems. From the statistical analysis of the results, we found that giant stars behave differently than subgiants and main-sequence stars, and that there is a good correlation between rotation periods and luminosities.
\end{abstract}

\begin{keywords}
stars: activity - stars: rotation - stars: binaries: spectroscopic.
\end{keywords}



\section{Introduction}
\label{S:intro}

RS CVn stars are close binaries where the hottest component, the \textit{primary}, is a subgiant or giant star of spectral type between F and K, and the coolest component, the \textit{secondary}, is a dwarf or subgiant of spectral type from G to M \citep{biermann1976}. The prototypical example, which gives its name to the group, is the eclipsing binary RS Canum Venaticorum. As a result of the much fainter luminosity of the coolest star, most of these systems are of the SB1 type.

These stars are fast rotators tidally synchronized, and are therefore more active than single stars with the same characteristics \citep{hall1972,Dempsey1993}, and they usually show strong H\(\alpha\), Ca \II\-, X-ray, and microwave emissions. This activity is dominated by the hottest component, which exhibits a high rotational rate \citep{weiler1978}. In the optical band, the most prominent feature is the periodic photometric variability, much larger than in individual stars with the same spectral type \citep{Dempsey1993}. This variability is due to the rotational modulation by large starspots, covering almost 10\% of the stellar surface \citep{Vogt1999}. 

\citet{walter1978} determined that RS CVn are also strong X-ray sources, with fluxes between $10^{26}$ and $10^{31.5}$ erg s\textsuperscript{-1}, and \citet{Dempsey1993} found that the temporal variations in X-rays are of the order of the starspot life-time. \citet{Singh1996} proposed that the X-ray variability is related to the tidally induced rotation that enhances the dynamo activity, since they did not found clear evidence that it might be due to mass transference between the components.

On the other hand, many of these systems show long-term chromospheric activity variations and activity cycles \citep{hall1991,Berdyugina07,messina08, Buccino2009, martinez2019}. To study these cycles is key to understand the connection between stellar rotation and long-term activity (see for example \citealt{rodono86}; \citealt{Berdyugina99}), and it is relevant to constrain the mechanism responsible for the generation and maintenance of the stellar magnetic fields \citep{weiler1978}. With this goal, we have studied the long-term activity of different main-sequence stars, and we have detected cycles in many of them \citep{2007A&A...461.1107C,2007A&A...474..345D,2011AJ....141...34B,2014ApJ...781L...9B,2017MNRAS.464.4299F,2019MNRAS.483.1159I,2019A&A...628L...1I}.

From statistical studies of activity in late-type stars with rotation periods between 22 and 48 days, \citet{noyes1984b} found a possible correlation between the activity-cycle period $P_\mathrm{cyc}$ and the Rossby number $Ro = P_\mathrm{rot} / \tau_\mathrm{c}$, where $\tau_\mathrm{c}$ is the convective turnover time near the bottom of the convection zone.
 
Later, \citet{saar1992} reported that stellar activity cycles form two distinct branches when plotted as a function of the rotation period, a fact that was later confirmed by \citet{Saar1999, bohm2007} and \citet{Brandenburg2017}. 
\citet{Brandenburg1998} found that the ratio $P_\mathrm{cyc}$/$P_\mathrm{rot}$ is proportional to the $\alpha$-effect in the mean field dynamo theory. \citet{Tuominen1988} found that this ratio decreases systematically with increasing thickness of the convection zone. 
Outside the main-sequence, evolved stars can develop increasingly thick convective envelopes and their convective turnover time increases \citep{Gilliland1985}. Simultaneously, the rotation period of the giant stars will also increase due to rotational braking \citep{Skumanich1972}. Therefore, studying activity cycles in evolved stars should help understand the dynamo mechanisms responsible for stellar activity.

In this work, we look for activity cycles on a large set of RS CVn-type stars, using photometric data obtained by the All Sky Automated Survey (ASAS, \citealt{Pojmanski2002}), and our own photometry obtained with the Optical Robotic Observatory (ORO). In Section 2 we present our stellar list and the observations dataset. In Section 3 we describe the method used to study the temporal evolution of the data, the correlation between the stellar variables that characterize our targets and the statistical analysis. Finally, in Section 4 we discuss the results.

\section{Observations and Data collection}

\subsection{Target list}

To our knowledge, there is not available a reliable list of RS CVn-type stars in the southern hemisphere. Therefore, it is important to build a list as complete as possible of confirmed RS CVn stars. In first approach, we obtained a preliminary classification from an exhaustive search in the bibliography of bright southern stars ($V \leq 14$ and $Dec<0$).

We first explored the southern brighter stars catalog of \citet{biedelman1973} together with the southern RS CVn candidate list of \citet{Weiler1979}. We confirm the targets as RS CVn-type using the catalogs of active binary stars published by \citet{rodono86}, \citet{Strassmeier1988}, \citet{Pallavicini1992} and \citet{Strassmeier1993}; the RS CVn binaries catalog of \citet{morris1988} for radio sources and \citet{Fleming1989} for X-ray sources; and the stellar classification catalogs given by \citet{Berdnikov2008} and \citet{drake2014}. We also included all the stars individually classified by \citet{pollard1989}, \citet{pandey2005}, \citet{Strassmeier2000}, \citet{Drake2006}, \citet{galvez2009} and \citet{lopez2012}. 

Finally, we added a few stars following the criteria established by \citet{Hall1976}\footnote{ This classification describes the main features of the RS CVn stars: spectral types between F-M with one or both evolved components, high rotation rate, chromospheric and X-ray emission.}.
The spectral type of the stars were obtained from the classification catalogs of \citet{houk1982}, \citet{Stocke1991}, \citet{sb92004}, \citet{torres2006} and \citet{skiff2014};
the catalogs of chromospherically active stars by \citet{fekel1986}, \citet{popper1991}, \citet{randich1993}, \citet{Thatcher1993}, \citet{fekel2005}, \citet{Bernhard2010} and \citet{saika2018}; and the of X-ray emission sources by \citet{Cutispoto1996}, \citet{kang1996}, \citet{cutispoto99}, \citet{hunsch1999}, \citet{Fuhrmeister2003}, \citet{Shtykovskiy2005}, \citet{Haakonsen2009} and \citet{Kiraga12}. 

In total, our final list, presented in Table \ref{tab:resumen1}, includes 121 sources. 


A statistic exploratory analysis allowed us to characterize our target list and estimate the distribution of different stellar parameters, also listed in Table \ref{tab:resumen1}. These quantities were obtained from the next catalogs:

\begin{itemize}
    \item The Hipparcos Catalogue \citep{perryman1997}: $V$ y $B$ magnitudes, $B-V$ color index.
    \item ASAS Photometry of ROSAT Sources \citep{Kiraga12}: $V$ magnitude.
    \item The Tycho-2 catalogue \citep{tycho2000}: $V$ and $B$ magnitudes, $B-V$ color index.
    \item The International Variable Star Index \citep{vsx2006}: rotation periods.
    \item The PASTEL catalogue, 2016 version \citep{pastel2016}: effective temperature $T_\mathrm{eff}$. 
    \item Gaia Data Release 2 \citep{gaia2018}: $G$ (330-1050 nm), $G_{BP}$ (330-680 nm) and $G_{RP}$ (630–1050 nm) apparent magnitudes, $T_\mathrm{eff}$ and luminosity $\log L$.
\end{itemize}

\begin{figure}
  \centering
  \includegraphics[width=\columnwidth]{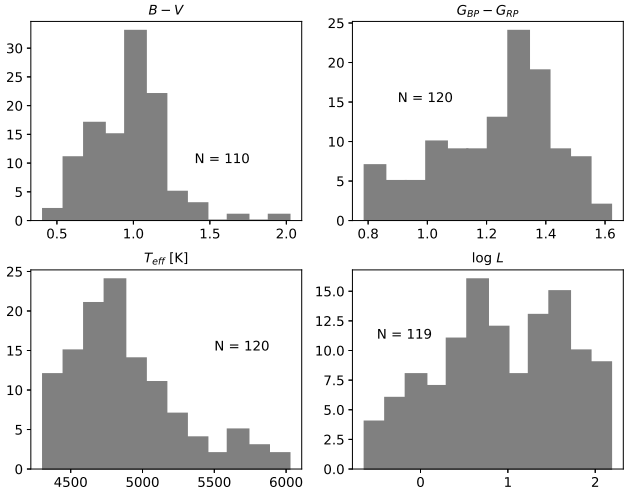}
  \caption{Distribution of the stellar parameters in our list. The shaded region shows the results for the whole sample, and the coloured lines the clustering groups described in Sect. 3}
  \label{fig:histo1}
 \end{figure}

In Fig. \ref{fig:histo1} we show the histograms for the colours $B-V$ and $G_{BP}-G_{RP}$, the effective temperature $T_\mathrm{eff}$ and the luminosity $\log L$. In each graph $N$ is the number of sources for which the specific quantity was available.

\subsection{Observations}

We used photometric data provided by the ASAS-3 database, which is available for our 121 stars. We only used observations qualified as A or B in the database (\textit{best} or \textit{mean} quality according to the ASAS definition). 

We also carried out photometric observations from February 2017 to January 2019 using the three 16'' MEADE telescopes that are part of the Optical Robotic Observatory (ORO) installed at the Estación en Altura Carlos U. Cesco, located in San Juan, Argentina. We observed 8-10 objects per night in non-consecutive nights to have a nearly uniform sample throughout their expected rotation periods. The image calibration and aperture differential photometry were optimally performed using our package \textit{killastro}, developed under the IRAF\footnote{The Image Reduction and Analysis Facility (IRAF) is distributed by the Association of Universities for Research in Astronomy (AURA), Inc., under contract to the National Science Foundation} platform. This package tries 10 different apertures, ranging from 0.7 to 4 FWHM, and chooses the aperture which minimizes the flux dispersion. We then performed differential photometry on these images, using as reference stars between 2 and 4 stars which do not show variability in the ASAS database. Finally, we intercalibrate with ASAS using the ASAS magnitude of the brightest reference star in the field 
(for details, see \citealt{martinez2019}). 

\section{Activity cycles analysis}

The Lomb–Scargle (LS) periodogram \citep{Horne1986} has been widely used to search for stellar activity cycles. \citet{Zechmeister2009} proposed a modification which has certain advantages in comparison to the classic LS periodogram: the Generalized Lomb–Scargle (GLS) periodogram. Namely, it takes into account a varying zero point, it does not require bootstrap or Monte Carlo algorithms to compute the significance of a signal, reducing the computational cost, and it is less susceptible to aliasing than the LS periodogram. In this study, we used the Python libraries developed by \citet{czesla2019}.

First, we applied to the light curves of all the sources the GLS periodogram in the 0.1 to 150 days range, to determine the rotation period of each target. This range was selected according to the maximum and minimum orbital periods of RS CVn-type systems found by \citet{Strassmeier1993}. However, we usually found weak peaks that could not be distinguished from noise, or with high levels of FAP. In fact, the photometric modulation attributed to rotation is usually not persistent throughout the entire data set. The absence of a well-defined photometric modulation at different epochs could be the result of longitudinal distributions of spots which are more or less homogeneous \citep{Murdoch1995}. Therefore, we looked in the periodogram for long-term variations associated with possible activity cycles, choosing periods longer than 150 days, and we obtained significant periods for 91 stars.


Then, we fitted the light curves with a sinusoidal function for the period found, we applied the GLS to the residuals for the period range between 0.1 and 150 days, and we found rotation periods for 102 sources. In several cases, marked with a $\bigstar$ in Table \ref{tab:resumen1}, we found that the amplitude of the rotational modulation changes during the activity cycle, due to the different filling factor of spots and active regions. 

We complemented this analysis with periods for 13 sources found in \textit{The International Variable Star Index} (VSX, \citealt{vsx2006}). For the eclipsing binaries, indicated with a $\dagger$ in Table \ref{tab:resumen1}, we used the phase dispersion minimization algorithm (PDM, \citealt{Stellingwerf1978}) provided in IRAF, and we list the orbital period instead. All the periods obtained are listed in Table \ref{tab:resumen1}, and the phase curves are shown in Figure \ref{fig:all_per1}. It can be seen in this figure that in a few cases the phase coverage of the observations in not adequate. We plan to complete the coverage with future observations. 

\begin{figure} \centering \includegraphics[width=\columnwidth]{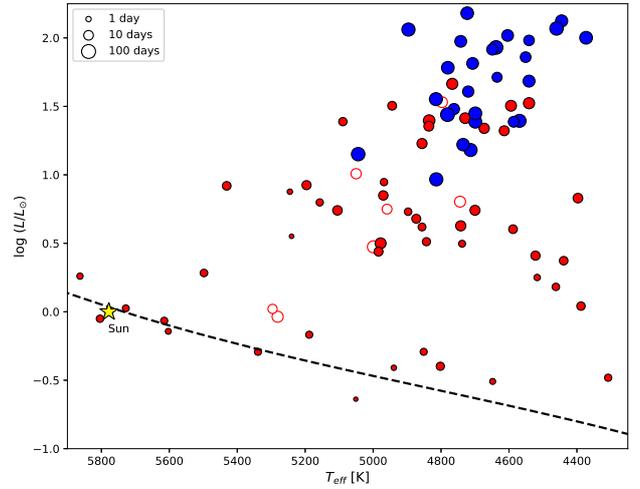}
\caption{HR diagram: Luminosity vs. effective temperature. The clustering we found is shown in red and blue. The size of each point is proportional to the rotation period. The theoretical main-sequence is represented by a black dotted line.}
\label{fig:HR} \end{figure}

We also applied to the data set the hierarchical agglomerative clustering with Ward linkage and Euclidean distance described by \citet{ward1963}, considering the variables $P_\mathrm{rot}$, $P_\mathrm{cyc}$, $A_\mathrm{cyc}$, $T_\mathrm{eff}$ and $\log L$. In fact, we found that the sample can be separated in two different clusters. In Fig. \ref{fig:HR} we show our stars in the HR diagram, with both clusters indicated with different colours. It can be seen that the stars marked in blue are giant stars, while those marked in red are closer to the main-sequence. 

\begin{figure}
  \centering
  \includegraphics[width=\columnwidth]{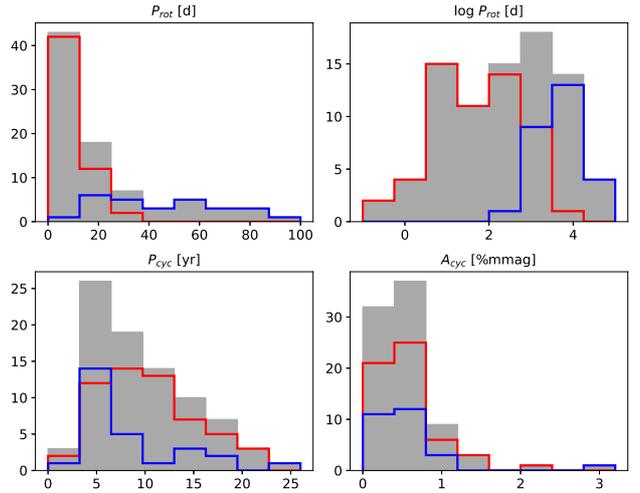}
  \caption{Distribution of the GLS results: $P_\mathrm{rot}$ in two different scales, $P_\mathrm{cyc}$ and its amplitude $A_\mathrm{cyc}$. The shaded regions show the results for the whole sample, and the coloured lines represent the clustering groups; there are 56 objects in cluster 1 (red) and 27 objects in cluster 2 (blue).}
  \label{fig:histoppr}
 \end{figure}

\begin{figure}\centering \includegraphics[width=\columnwidth]{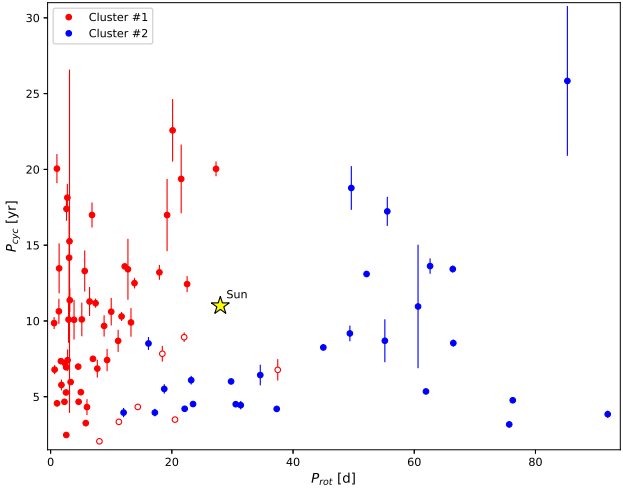} \caption{Rotation vs. activity periods. The two groups found in the clustering are shown in red and blue. The position of the Sun is indicated by the yellow star.} \label{figs:pp_cluster} \end{figure}

In Fig. \ref{fig:histoppr} we show the distribution of our results for the whole sample and for both clusters. We plot $P_\mathrm{rot}$, $P_\mathrm{cyc}$ and the activity cycle amplitude $A_\mathrm{cyc}$, expressed as percentage of the mean magnitude $V$ (\% mmag). It can be seen that the red group has shorter rotation periods and cycles both longer and with larger amplitudes. This last result can also be seen in Fig. \ref{figs:pp_cluster}, where we plot the activity period $P_\mathrm{cyc}$ as a function of the rotation period $P_\mathrm{rot}$ for each agglomerate. 

For the cluster \#2 (blue color), associated to giant stars, we found that ${P_\mathrm{cyc}}/ {P_\mathrm{rot}} = 75 \pm 38$ and $\log ({P_\mathrm{cyc}}/ {P_\mathrm{rot}}) = 1.81 \pm 0.26 $. For comparison, \citet{bohm2007}, in a study of main-sequence stars, obtained $\log ({P_\mathrm{cyc}}/ {P_\mathrm{rot}}) = 1.95 $ for the {\it{inactive}} branch.  



 Finally, to look for correlations between the stellar properties in our target list, we used the non-parametrical Spearman correlation coefficient $r_{ij}$ \citep{conover1999}, which can be applied to variables with non-Gaussian distributions. The quantities we analyzed form a variable vector $\mathbf{v}$, given by:
\begin{equation}
\mathbf{v} =
\left[
 \begin{matrix}
P_\mathrm{rot}	&	P_\mathrm{cyc}	&	A_\mathrm{cyc}	& T_\mathrm{eff}	& \log L	\\
 \end{matrix}
 \right],
 \label{Eq:vector}
\end{equation}
and the results can be summarized by a matrix $\mathbf{R}$. In this matrix, the elements $r_{ij}$ below the main diagonal represent the correlation between the \textit{i}-th y \textit{j}-th variables, and the elements $p_{ij}$ above the main diagonal represent the probability that the \textit{i}-th and \textit{j}-th variables are independent.

The $r_{ij}$ coefficients indicate if the correlation is perfect ($\abs{1}$), excellent ($\abs{0.9-1}$), acceptable ($\abs{0.75-0.9}$), regular ($\abs{0.5-0.75}$), weak ($\abs{0.25-0.5}$) or absent ($<0.25$). $r_{ij}$ is positive for a direct correlation and negative for an inverse one. 

The result we obtained is: 
\begin{equation}
\mathbf{R} =
\left[
 \begin{matrix}
        1	&	0.64	&	0.1	&	\ast	&	\ast	\\
    -0.05	&	1	&	\ast	&	0.51	&	0.59	\\
    0.18	&	0.63	&	1	&	0.01	&	0.83	\\
    -0.36	&	-0.07	&	-0.26	&	1	&	\ast	\\
     0.76	&	-0.06	&	0.02	&	-0.35	&	1	\\
 \end{matrix}
 \right].
\end{equation}

The matrix elements pointed out with a $\ast$ represent probability values lower than 0.0005.
From these results, it seems that there is a good positive correlation between rotation period $P_\mathrm{rot}$ and luminosity $\log L$, and a regular correlation between the activity period $P_\mathrm{cyc}$ and its amplitude $A_\mathrm{cyc}$.

\begin{figure} \centering
\includegraphics[width=\columnwidth]{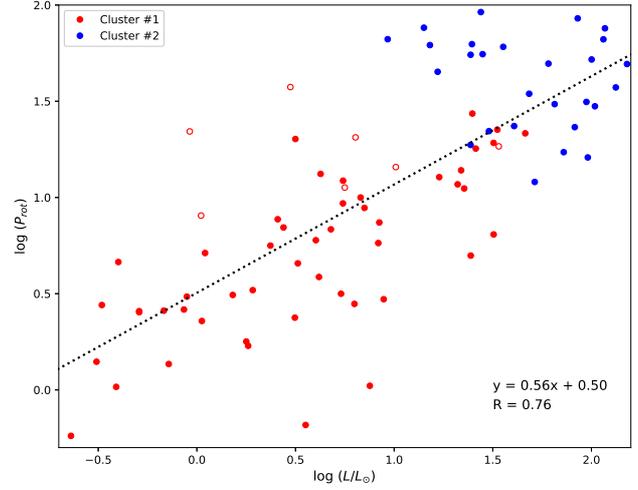}
\caption{$\log (P_\mathrm{rot})$ versus $\log (L/L_\odot)$. The clustering is shown with the same colour code than in the other figures. The $R$-spearman coefficient for this correlation is 0.76.}
\label{fig:ppl1} \end{figure}

We therefore plot the rotation periods $P_\mathrm{rot}$ in logarithmic scale as a function of the luminosity $L/L_\odot$ in Fig.  \ref{fig:ppl1}. The correlation found in (2), with $R_{s}=0.76$, is apparent. The two groups seen in Fig. \ref{fig:HR} are separated by luminosity. Fitting this correlation, we obtain that:
\begin{equation}
    P_\mathrm{rot} \approx 
    3.16\ \left( \frac{L}{L_\odot} \right)^{3/5} . 
\end{equation}



\section{Conclusions}

In this work we present a  list of 121 bona fide RS CVn southern stars  with $V \leq 14$. For this list, we compiled the main stellar parameters: the colour index $B-V$, the effective temperature $T_\mathrm{eff}$ and the luminosity $L$ (see \ref{tab:resumen1})

We also analyzed the activity of these systems in different time scales. Applying the Generalized Lomb-Scargle periodogram (GLS) to photometric data obtained from the publicly available ASAS-3 database and our own observations obtained with ORO, we found rotation periods for 102 systems, which we completed with results from the literature for 13 more. We also obtained activity cycles for 91 systems. In many cases, we observed that the amplitude variation due to the rotation is modulated by the activity cycle, due to the different area covered by starspots. 

We applied a hierarchical agglomerative clustering method to our data and found that giant stars are separated from subgiants and main-sequence stars. 

We also found a good correlation between the rotation period $P_\mathrm{rot}$ and the luminosity $\log L$, and a fair one between the activity period $P_\mathrm{cyc}$ and the percent amplitude variation $A_\mathrm{cyc}$.

\section{Acknowledgments}

We are grateful to the personal of the Observatorio Astronómico Félix Aguilar for their invaluable help with the operation of the Optical Robotic Observatory.

\section{Data availability}

The data underlying this article will be shared on reasonable request to the corresponding author.



\bibliographystyle{mnras}
\bibliography{mybiblio} 


\begin{figure*}
    \centering
    \caption{Phase curves for the  activity cycles we detected ($P$ is expressed in years).}
    \includegraphics[width=\textwidth]{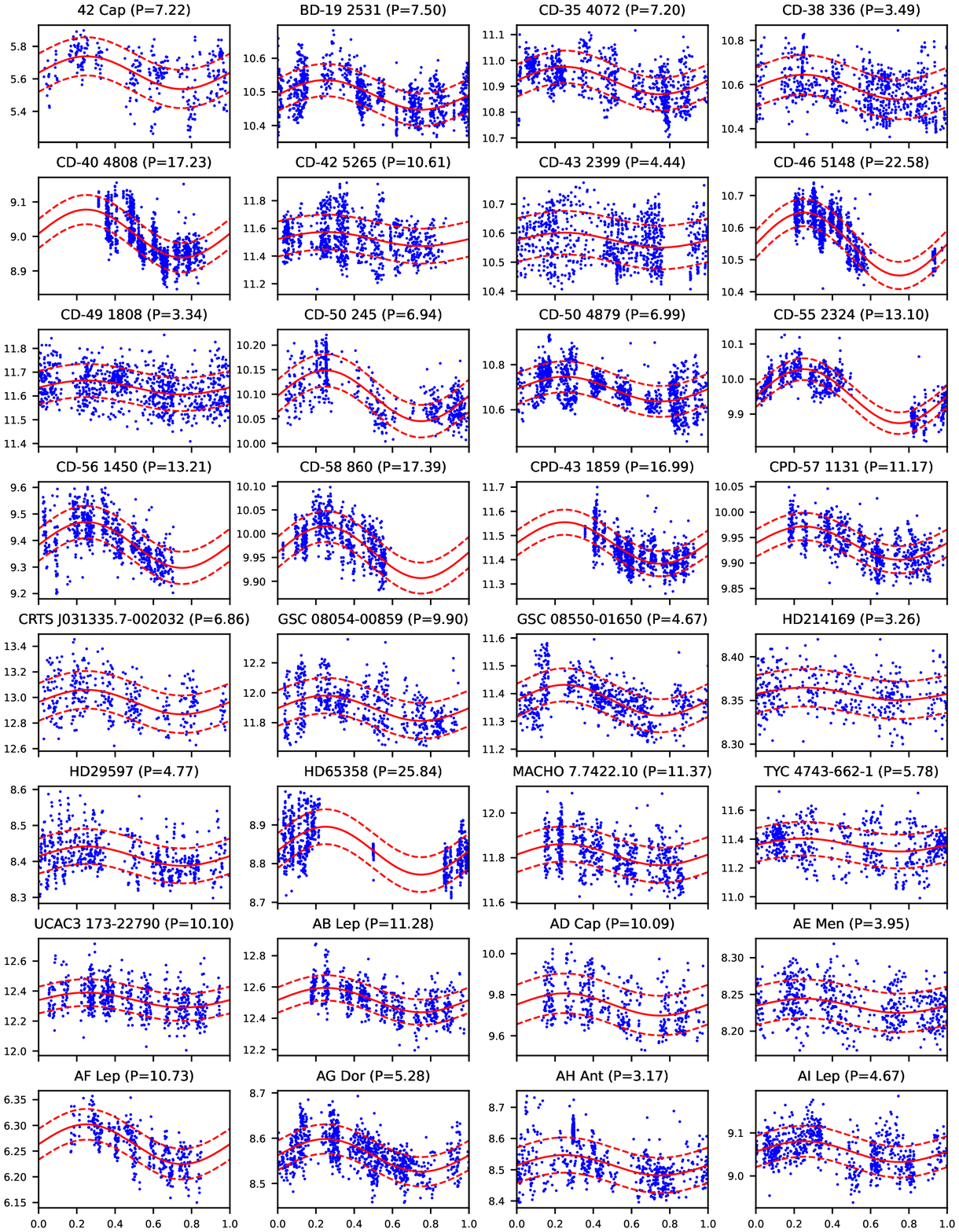}
    \label{fig:all_per1}
\end{figure*}

\begin{figure*}
    \centering
    \caption*{\textit{(continued)}}
    \includegraphics[width=\textwidth]{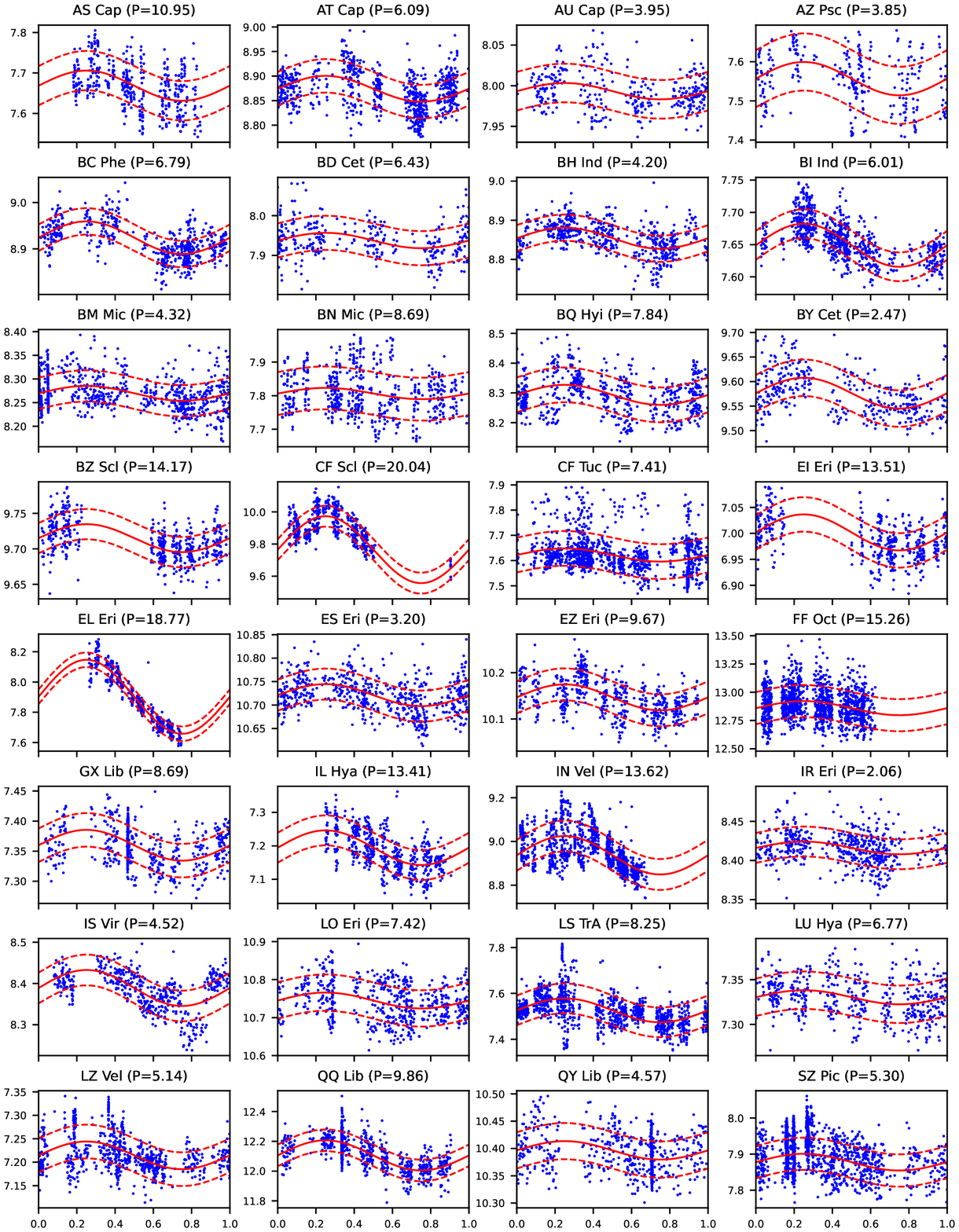}
\end{figure*}

\begin{figure*}
    \centering
    \caption*{\textit{(continued)}.}
    \includegraphics[width=\textwidth]{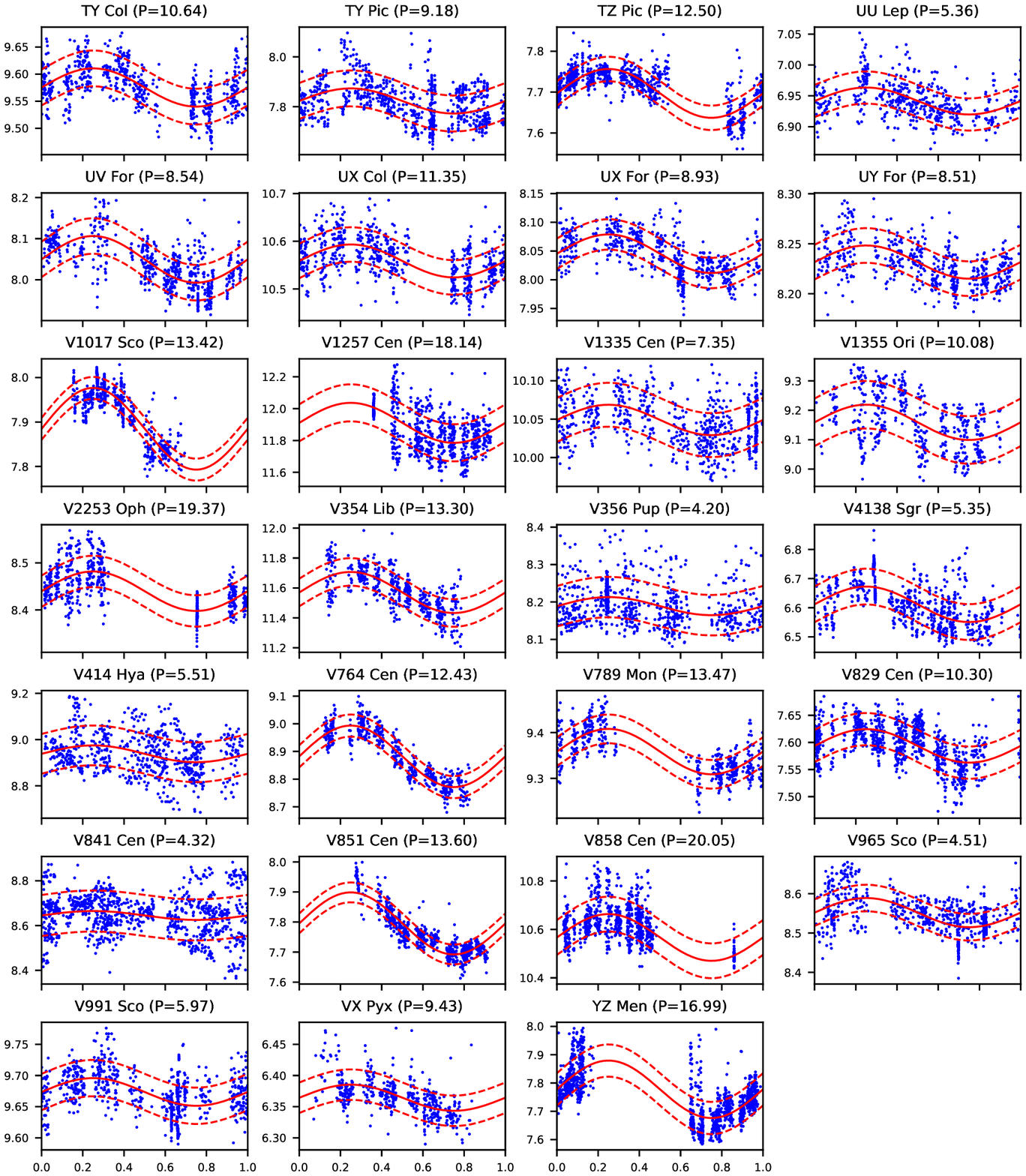}
\end{figure*}

\begin{table*}
\centering
\caption{The complete list of our targets, including our results.
The different columns list (from left to right) the star ID, the $V$ magnitude, the $B-V$ colour index, the effective temperature of the binary system, the luminosity in logarithm scale, the short-period associated to the stellar rotation, 
the error in the short-period determination, the FAP corresponding to the short-period, the long-period associated to the stellar activity cycle, 
the long-period amplitude, the error in the long-period determination and the FAP corresponding to the long-period. $\maltese$ rotation period obtained from \textit{The International Variable Star Index (VSX)}. $\bigstar$ stars having a rotation amplitude which varies during the activity cycle. $\dagger$ eclipsing binaries. }
\begin{adjustbox}{width=\textwidth}
\begin{tabular}{lccccccccc}
\hline\hline\noalign{\smallskip}
    ID & $V$ & $B-V$ & $T_\mathrm{eff}$ & $\log L$    & $P_\mathrm{rot}$  & 
    FAP ($P_\mathrm{rot}$) &  $P_\mathrm{cyc}$  & $A_\mathrm{cyc}$ & 
    FAP ($P_\mathrm{cyc}$) \\
       &     &       &  [K]             & [$L_{\odot}$] &      [d]                 &     &  [yr]             &  [\%]       &      \\
    \hline\noalign{\smallskip}
*42 Cap 	  	&	5.18	&	0.65	&	5682	&	0.99	&	-			&		 - 		&$	7.22	\pm	2.39	$&	1.79	&	$	10^{-4	}$	\\
*del Eri 		&	3.54	&	0.92	&	5023	&	0.60	&	-			&		-		&	-			&	-	&		-		\\
1E2157.6-1016		&	12.54	&	-	&	4395	&	0.36	&$	7.108	\pm	0.001	$&		-		&	-			&	-	&		-		\\
BD-19 2531 	  	&	10.51	&	0.69	&	4984	&	0.44	&$	6.984	\pm	0.001	$&	$	<10^{-15	}$	&$	7.50	\pm	0.16	$&	0.42	&	$	<10^{-15	}$	\\
CD-35 4072 	  	&	10.93	&	0.83	&	4738	&	0.50	&$	2.373	\pm	0.0001	$&	$	<10^{-15	}$	&$	7.20	\pm	0.18	$&	0.49	&	$	<10^{-15	}$	\\
CD-38 336 	 $\bigstar$ 	&	10.59	&	1.26	&	4744	&	0.80	&$	20.521	\pm	0.006	$&	$	<10^{-15	}$	&$	3.49	\pm	0.03	$&	0.53	&	$	<10^{-15	}$	\\
CD-40 4808 	 $\bigstar$ 	&	9.00	&	1.13	&	4699	&	1.45	&$	55.552	\pm	0.055	$&	$	<10^{-15	}$	&$	17.23	\pm	0.96	$&	0.77	&	$	<10^{-15	}$	\\
CD-42 130		&	10.55	&	1.68	&	4989	&	0.72	&$	16.941	\pm	0.007	$&		-		&	-			&	-	&		-		\\
CD-42 5265 	 $\bigstar$ 	&	11.44	&	1.16	&	4397	&	0.83	&$	9.998	\pm	0.002	$&	$	<10^{-15	}$	&$	10.61	\pm	0.91	$&	0.44	&	$	<10^{-15	}$	\\
CD-43 2399 	 $\bigstar$ 	&	10.58	&	0.83	&	4742	&	1.97	&$	31.369	\pm	0.007	$&	$	<10^{-15	}$	&$	4.44	\pm	0.28	$&	0.24	&	$	0.08	$	\\
CD-46 5148 	  	&	10.40	&	0.93	&	4978	&	0.50	&$	20.127	\pm	0.006	$&	$	<10^{-15	}$	&$	22.58	\pm	2.06	$&	0.93	&	$	10^{-14	}$	\\
CD-49 1808 	  	&	11.67	&	1.09	&	4959	&	0.75	&$	11.259	\pm	0.002	$&	$	<10^{-15	}$	&$	3.34	\pm	0.07	$&	0.25	&	$	<10^{-15	}$	\\
CD-50 245 	  	&	10.11	&	0.77	&	5339	&	-0.30	&$	2.566	\pm	0.0002	$&	$	<10^{-15	}$	&$	6.94	\pm	0.08	$&	0.51	&	$	<10^{-15	}$	\\
CD-50 2604		&	9.74	&	0.63	&	5336	&	1.70	&$	24.289	\pm	0.011	$&		-		&	-			&	-	&		-		\\
CD-50 4879 	 $\bigstar$ 	&	10.94	&	0.98	&	4843	&	0.51	&$	4.547	\pm	0.0003	$&	$	<10^{-15	}$	&$	6.99	\pm	0.14	$&	0.51	&	$	<10^{-15	}$	\\
CD-55 2324 	  	&	10.00	&	2.02	&	4373	&	2.00	&$	52.133	\pm	0.050	$&	$	<10^{-15	}$	&$	13.10	\pm	0.20	$&	0.78	&	$	<10^{-15	}$	\\
CD-56 1450 	 $\bigstar$ 	&	9.55	&	0.97	&	4729	&	1.41	&$	17.948	\pm	0.019	$&	$	10^{-7	}$	&$	13.21	\pm	0.49	$&	0.92	&	$	<10^{-15	}$	\\
CD-58 860 	  	&	10.01	&	0.80	&	5615	&	-0.07	&$	2.615	\pm	0.0002	$&	$	<10^{-15	}$	&$	17.39	\pm	0.78	$&	0.55	&	$	<10^{-15	}$	\\
CPD-43 1859 	  	&	11.45	&	0.98	&	4873	&	0.68	&$	6.830	\pm	0.001	$&	$	<10^{-15	}$	&$	16.99	\pm	0.82	$&	0.75	&	$	<10^{-15	}$	\\
CPD-57 1131 	  	&	9.95	&	0.75	&	5196	&	0.92	&$	7.414	\pm	0.001	$&	$	<10^{-15	}$	&$	11.17	\pm	0.30	$&	0.33	&	$	<10^{-15	}$	\\
CRTS J031335.7-002032 	  	&	12.70	&	-	&	4522	&	0.41	&$	7.704	\pm	0.001	$&	$	<10^{-15	}$	&$	6.86	\pm	0.59	$&	0.74	&	$	10^{-6	}$	\\
CRTS J205535.2-153510		&	12.88	&	-	&	4760	&	1.08	&$	10.650	\pm	0.004	$&		-		&	-			&	-	&		-		\\
CRTS J225004.4-075247		&	13.16	&	-	&	4897	&	0.96	&$	42.112	\pm	0.054	$&		-		&	-			&	-	&		-		\\
GSC 08054-00859 	 $\bigstar$ 	&	11.71	&	1.09	&	4742	&	0.63	&$	13.257	\pm	0.002	$&	$	<10^{-15	}$	&$	9.90	\pm	0.94	$&	0.71	&	$	10^{-10	}$	\\
GSC 08550-01650 	  	&	11.43	&	0.90	&	4802	&	-0.39	&$	4.618	\pm	0.001	$&	$	10^{-9	}$	&$	4.67	\pm	0.10	$&	0.49	&	$	<10^{-15	}$	\\
Haro 5-72		&	13.86	&	-	&	4310	&	-0.25	&$	5.929	\pm	0.002	$&		-		&	-			&	-	&		-		\\
HD 214169 	  	&	8.38	&	0.96	&	5431	&	0.92	&$	5.796	\pm	0.001	$&	$	10^{-9	}$	&$	3.26	\pm	0.21	$&	0.09	&	$	0.001	$	\\
HD 29597 	 $\bigstar$ 	&	8.40	&	0.82	&	5044	&	1.15	&$	76.253	\pm	0.084	$&	$	<10^{-15	}$	&$	4.77	\pm	0.15	$&	0.32	&	$	<10^{-15	}$	\\
HD 65358 	 $\bigstar$ 	&	8.85	&	1.40	&	4638	&	1.93	&$	85.266	\pm	0.139	$&	$	<10^{-15	}$	&$	25.84	\pm	4.94	$&	0.70	&	$	<10^{-15	}$	\\
HD 155555		&	6.72	&	0.78	&	5129	&	0.28	&$	1.687	\pm	0.0002	$&		-		&	-			&	-	&		-		\\
HD 17925		&	6.05	&	0.86	&	5235	&	-0.39	&$	6.710			$&		-		&	-			&	-	&		-		\\
MACHO 7.7422.10 	  	&	11.88	&	-	&	4897	&	0.73	&$	3.161	\pm	0.0001	$&	$	<10^{-15	}$	&$	11.37	\pm	0.77	$&	0.40	&	$	<10^{-15	}$	\\
TYC 4743-662-1 	 $\bigstar$ 	&	11.25	&	1.44	&	4517	&	0.25	&$	1.784	\pm	0.00004	$&	$	<10^{-15	}$	&$	5.78	\pm	0.36	$&	0.40	&	$	10^{-8	}$	\\
TYC 8380-1953-1		&	10.45	&	1.06	&	4517	&	1.21	&$	3.076	\pm	0.0002	$&		-		&	-			&	-	&		-		\\
UCAC3 173-22790 	  	&	12.80	&	-	&	4388	&	0.04	&$	5.144	\pm	0.0005	$&	$	<10^{-15	}$	&$	10.10	\pm	1.11	$&	0.41	&	$	10^{-8	}$	\\
V* AB Lep 	  	&		&	-	&	4944	&	1.50	&$	6.425	\pm	0.001	$&	$	10^{-13	}$	&$	11.28	\pm	0.95	$&	0.62	&	$	<10^{-15	}$	\\
V* AD Cap 	 $\maltese$  $\dagger$ 	&	9.77	&	1.00	&	4968	&	0.95	&$	2.959	\pm	0.089	$&		 - 		&$	10.09	\pm	1.53	$&	0.56	&	$	10^{-15	}$	\\
V* AE Men 	  	&	8.26	&	1.11	&	4635	&	1.71	&$	12.038	\pm	0.002	$&	$	<10^{-15	}$	&$	3.95	\pm	0.31	$&	0.12	&	$	0.001	$	\\
V* AF Lep 	 $\dagger$ 	&	6.30	&	0.54	&	6496	&	0.25	&$	1.000			$&		 - 		&$	10.73	\pm	0.60	$&	0.62	&	$	<10^{-15	}$	\\
V* AG Dor 	 $\bigstar$ 	&	8.60	&	0.90	&	4851	&	-0.29	&$	2.534	\pm	0.0001	$&	$	<10^{-15	}$	&$	5.28	\pm	0.09	$&	0.42	&	$	<10^{-15	}$	\\
V* AH Ant 	  	&	8.40	&	1.20	&	4460	&	2.07	&$	75.671	\pm	0.104	$&	$	<10^{-15	}$	&$	3.17	\pm	0.20	$&	0.38	&	$	0.2	$	\\
V* AI Lep 	  	&	8.95	&	0.63	&	5728	&	0.03	&$	2.281	\pm	0.0002	$&	$	10^{-7	}$	&$	4.67	\pm	0.12	$&	0.27	&	$	<10^{-15	}$	\\
V* AK Vol		&	8.60	&	0.99	&	4739	&	1.83	&$	31.792	\pm	0.017	$&		-		&	-			&	-	&		-		\\
V* AR Mon	 $\dagger$ 	&	8.79	&	1.05	&	4709	&	1.84	&$	21.211			$&		-		&	-			&	-	&		-		\\
V* AS Cap 	 $\bigstar$ 	&	7.69	&	1.06	&	4815	&	1.55	&$	60.625	\pm	0.050	$&	$	<10^{-15	}$	&$	10.95	\pm	4.07	$&	0.49	&	$	0.02	$	\\
V* AT Cap 	 $\maltese$  $\dagger$ 	&	8.87	&	1.28	&	4649	&	1.92	&$	23.193			$&		-		&$	6.09	\pm	0.29	$&	0.30	&	$	<10^{-15	}$	\\
V* AU Cap 	  	&	8.01	&	1.21	&	4551	&	1.86	&$	17.196	\pm	0.012	$&	$	10^{-12	}$	&$	3.95	\pm	0.25	$&	0.13	&	$	10^{-5	}$	\\
V* AZ Psc 	 $\bigstar$ 	&		&	-	&	4781	&	1.44	&$	91.920	\pm	0.221	$&	$	<10^{-15	}$	&$	3.85	\pm	0.24	$&	0.56	&	$	0.01	$	\\
V* BC Phe 	 $\dagger$ 	&	8.92	&	0.71	&	5240	&	0.55	&$	0.657			$&		-		&$	6.79	\pm	0.30	$&	0.39	&	$	<10^{-15	}$	\\
V* BD Cet 	 $\bigstar$ 	&	7.98	&	1.11	&	4541	&	1.68	&$	34.593	\pm	0.026	$&	$	10^{-14	}$	&$	6.43	\pm	0.68	$&	0.24	&	$	10^{-6	}$	\\
V* BH Ind 	  	&	9.12	&	1.04	&	4762	&	1.48	&$	22.117	\pm	0.020	$&	$	10^{-7	}$	&$	4.20	\pm	0.09	$&	0.30	&	$	<10^{-15	}$	\\
\end{tabular}
\end{adjustbox}
\centering
\label{tab:resumen1}
\end{table*}

\begin{table*}
\centering
    \caption*{\textit{(continued)}}
    \begin{adjustbox}{width=\textwidth}
    \begin{tabular}{lccccccccc}
    \hline\noalign{\smallskip}
    ID & $V$ & $B-V$ & $T_\mathrm{eff}$ & $\log L$    & $P_\mathrm{rot}$  & 
    FAP ($P_\mathrm{rot}$) &  $P_\mathrm{cyc}$  & $A_\mathrm{cyc}$ & 
    FAP ($P_\mathrm{cyc}$) \\
       &     &       &  [K]             & [$L_{\odot}$] &      [d]                 &     &  [yr]             &  [\%]       &      \\
    \hline\noalign{\smallskip}
V* BH Vir	 $\dagger$ 	&	9.68	&	0.61	&	5876	&	0.43	&$	0.817			$&		-		&	-			&	-	&		-		\\
V* BI Ind 	  	&	7.67	&	1.22	&	4604	&	2.02	&$	29.775	\pm	0.017	$&	$	<10^{-15	}$	&$	6.01	\pm	0.10	$&	0.44	&	$	<10^{-15	}$	\\
V* BM Mic 	 $\bigstar$ 	&	8.29	&	0.77	&	5050	&	1.01	&$	14.382	\pm	0.007	$&	$	<10^{-15	}$	&$	4.32	\pm	0.08	$&	0.19	&	$	<10^{-15	}$	\\
V* BN Mic 	  	&	7.76	&	1.11	&	4699	&	1.39	&$	55.149	\pm	0.047	$&	$	<10^{-15	}$	&$	8.69	\pm	1.42	$&	0.22	&	$	10^{-4	}$	\\
V* BQ Hyi 	  	&	9.12	&	0.72	&	4797	&	1.53	&$	18.430	\pm	0.005	$&	$	<10^{-15	}$	&$	7.84	\pm	0.52	$&	0.41	&	$	<10^{-15	}$	\\
V* BY Cet 	 $\bigstar$ 	&	9.60	&	0.77	&	5188	&	-0.17	&$	2.580	\pm	0.0002	$&	$	10^{-12	}$	&$	2.47	\pm	0.05	$&	0.33	&	$	<10^{-15	}$	\\
V* BZ Scl 	  	&	9.68	&	0.64	&	5804	&	-0.05	&$	3.053	\pm	0.0002	$&	$	10^{-14	}$	&$	14.17	\pm	3.16	$&	0.20	&	$	<10^{-15	}$	\\
V* CF Oct		&	7.93	&	1.07	&	4881	&	1.62	&$	20.376	\pm	0.010	$&		-		&	-			&	-	&		-		\\
V* CF Scl 	  	&	9.93	&	0.70	&	4835	&	1.40	&$	27.292	\pm	0.012	$&	$	<10^{-15	}$	&$	20.04	\pm	0.49	$&	2.13	&	$	<10^{-15	}$	\\
V* CF Tuc 	 $\maltese$  $\dagger$ 	&	7.90	&	0.41	&	5157	&	0.80	&$	2.798			$&		-		&$	7.41	\pm	0.71	$&	0.34	&	$	<10^{-15	}$	\\
V* EI Eri 	  	&	7.04	&	0.67	&	5513	&	0.59	&	-			&		-		&$	13.51	\pm	4.31	$&	0.50	&	$	<10^{-15	}$	\\
V* EL Eri 	 $\bigstar$ 	&	8.19	&	1.10	&	4780	&	1.78	&$	49.614	\pm	0.065	$&	$	10^{-15	}$	&$	18.77	\pm	1.44	$&	3.09	&	$	<10^{-15	}$	\\
V* ER Eri		&	9.85	&	1.10	&	4666	&	0.69	&$	5.920	\pm	0.001	$&		-		&	-			&	-	&		-		\\
V* ES Eri 	  	&	10.63	&	0.98	&	5078	&	-0.18	&	-			&		-		&$	3.20	\pm	0.06	$&	0.22	&	$	<10^{-15	}$	\\
V* EZ Eri 	  	&	10.17	&	1.03	&	4970	&	0.85	&$	8.824	\pm	0.001	$&	$	<10^{-15	}$	&$	9.67	\pm	0.70	$&	0.28	&	$	<10^{-15	}$	\\
V* FF Oct 	  	&		&	-	&	4462	&	0.18	&$	3.114	\pm	0.0002	$&	$	<10^{-15	}$	&$	15.26	\pm	11.32	$&	0.49	&	$	10^{-5	}$	\\
V* GX Lib 	  	&	7.35	&	1.04	&	4836	&	1.36	&$	11.139	\pm	0.001	$&	$	<10^{-15	}$	&$	8.69	\pm	0.73	$&	0.35	&	$	<10^{-15	}$	\\
V* HU Vir		&	8.71	&	0.98	&	4517	&	0.74	&$	10.438	\pm	0.002	$&		-		&	-			&	-	&		-		\\
V* IL Hya 	 $\bigstar$ 	&	7.37	&	1.01	&	4856	&	1.23	&$	12.755	\pm	0.003	$&	$	<10^{-15	}$	&$	13.41	\pm	2.01	$&	0.72	&	$	<10^{-15	}$	\\
V* IN Vel 	 $\bigstar$ 	&	9.08	&	1.18	&	4569	&	1.39	&$	62.611	\pm	0.068	$&	$	<10^{-15	}$	&$	13.62	\pm	0.51	$&	0.99	&	$	<10^{-15	}$	\\
V* IN Vir		&	9.13	&	1.20	&	4436	&	0.75	&$	8.129	\pm	0.001	$&		-		&	-			&	-	&		-		\\
V* IR Eri 	 $\dagger$ 	&	8.45	&	0.85	&	5296	&	0.02	&$	8.050			$&		-		&$	2.06	\pm	0.05	$&	0.10	&	$	10^{-8	}$	\\
V* IS Vir 	 $\dagger$ 	&	8.43	&	1.07	&	4720	&	1.61	&$	23.500			$&		-		&$	4.52	\pm	0.10	$&	0.52	&	$	<10^{-15	}$	\\
V* LO Eri 	 $\bigstar$ 	&	10.76	&	0.89	&	5105	&	0.74	&$	9.318	\pm	0.001	$&	$	<10^{-15	}$	&$	7.42	\pm	0.73	$&	0.19	&	$	10^{-10	}$	\\
V* LS TrA 	 $\bigstar$ 	&	8.14	&	1.04	&	4735	&	1.22	&$	44.997	\pm	0.041	$&	$	<10^{-15	}$	&$	8.25	\pm	0.20	$&	0.69	&	$	<10^{-15	}$	\\
V* LU Hya 	  	&	7.35	&	0.95	&	4999	&	0.47	&$	37.474	\pm	0.050	$&	$	10^{-10	}$	&$	6.77	\pm	0.71	$&	0.11	&	$	10^{-6	}$	\\
V* LZ Vel 	 $\bigstar$ 	&	7.19	&	1.27	&	4420	&	3.39	&$	92.791	\pm	0.306	$&	$	<10^{-15	}$	&$	5.14	\pm	0.05	$&	0.41	&	$	<10^{-15	}$	\\
V* NR Lib		&	11.34	&	0.93	&	4571	&	0.30	&$	2.683	\pm	0.0001	$&		-		&	-			&	-	&		-		\\
V* QQ Lib 	 $\dagger$ 	&	11.89	&	0.56	&	5051	&	-0.64	&$	0.577			$&		-		&$	9.86	\pm	0.40	$&	0.84	&	$	<10^{-15	}$	\\
V* QY Lib 	  	&	10.41	&	0.81	&	5245	&	0.88	&$	1.050	\pm	0.00002	$&	$	<10^{-15	}$	&$	4.57	\pm	0.21	$&	0.16	&	$	<10^{-15	}$	\\
V* RV Lib	 $\dagger$ 	&	9.17	&	1.09	&	4708	&	1.41	&$	10.721			$&		-		&	-			&	-	&		-		\\
V* RZ Eri	 $\dagger$ 	&	7.88	&	0.65	&	5324	&	1.49	&$	39.282			$&		-		&	-			&	-	&		-		\\
V* SZ Pic 	  	&	7.91	&	0.75	&	5089	&	1.39	&$	4.988	\pm	0.001	$&	$	<10^{-15	}$	&$	5.30	\pm	0.17	$&	0.30	&	$	10^{-4	}$	\\
V* TW Lep		&	7.47	&	1.05	&	4805	&	1.71	&$	27.752	\pm	0.025	$&		-		&	-			&	-	&		-		\\
V* TX Pic		&	6.10	&	1.16	&	4559	&	2.13	&$	20.914	\pm	0.033	$&		-		&	-			&	-	&		-		\\
V* TY Col 	  	&	9.60	&	0.68	&	5603	&	-0.14	&$	1.363	\pm	0.0001	$&	$	10^{-6	}$	&$	10.64	\pm	0.83	$&	0.37	&	$	<10^{-15	}$	\\
V* TY Pic 	 $\bigstar$ 	&	7.70	&	0.99	&	4723	&	2.18	&$	49.378	\pm	0.037	$&	$	<10^{-15	}$	&$	9.18	\pm	0.51	$&	0.65	&	$	<10^{-15	}$	\\
V* TY Pyx	 $\dagger$ 	&	6.87	&	0.69	&	5669	&	0.67	&$	3.199			$&		-		&	-			&	-	&		-		\\
V* TZ Col		&	9.08	&	0.59	&	5907	&	0.26	&$	2.843	\pm	0.0003	$&		-		&	-			&	-	&		-		\\
V* TZ Pic 	  	&	7.74	&	1.05	&	4673	&	1.34	&$	13.839	\pm	0.008	$&	$	10^{-5	}$	&$	12.50	\pm	0.35	$&	0.78	&	$	<10^{-15	}$	\\
V* UU Lep 	  	&	6.94	&	1.11	&	4674	&	1.88	&	-			&		-		&$	5.36	\pm	0.21	$&	0.31	&	$	<10^{-15	}$	\\
V* UV For 	 $\bigstar$ 	&	8.06	&	0.99	&	4814	&	0.97	&$	66.447	\pm	0.135	$&	$	10^{-11	}$	&$	8.54	\pm	0.23	$&	0.71	&	$	<10^{-15	}$	\\
V* UX Col 	  	&	10.59	&	1.03	&	5076	&	-	&$	0.970	\pm	0.00002	$&	$	<10^{-15	}$	&$	11.35	\pm	1.20	$&	0.33	&	$	<10^{-15	}$	\\
V* UX For 	  	&	8.06	&	0.72	&	5281	&	-0.04	&$	22.035	\pm	0.016	$&	$	10^{-7	}$	&$	8.93	\pm	0.31	$&	0.42	&	$	<10^{-15	}$	\\
V* UY For 	 $\bigstar$ 	&	8.25	&	1.15	&	4541	&	1.98	&$	16.144	\pm	0.010	$&	$	10^{-9	}$	&$	8.51	\pm	0.42	$&	0.20	&	$	<10^{-15	}$	\\
V* V1017 Sco 	 $\bigstar$ 	&	7.99	&	1.16	&	4896	&	2.06	&$	66.356	\pm	0.152	$&	$	<10^{-15	}$	&$	13.42	\pm	0.23	$&	1.17	&	$	<10^{-15	}$	\\
V* V1257 Cen 	 $\bigstar$ 	&	11.68	&	1.01	&	4308	&	-0.48	&$	2.761	\pm	0.0003	$&	$	<10^{-15	}$	&$	18.14	\pm	0.91	$&	1.05	&	$	<10^{-15	}$	\\
V* V1265 Cen		&	11.01	&	-	&	4361	&	0.86	&$	3.533	\pm	0.0001	$&		-		&	-			&	-	&		-		\\
V* V1335 Cen 	  	&	10.06	&	0.59	&	5863	&	0.26	&$	1.695	\pm	0.0001	$&	$	10^{-11	}$	&$	7.35	\pm	0.16	$&	0.20	&	$	<10^{-15	}$	\\
V* V1355 Ori 	 $\bigstar$ 	&	9.19	&	0.86	&	4856	&	0.62	&$	3.863	\pm	0.0002	$&	$	<10^{-15	}$	&$	10.08	\pm	1.30	$&	0.66	&	$	<10^{-15	}$	\\
V* V2253 Oph 	 $\bigstar$ 	&	8.49	&	1.17	&	4767	&	1.66	&$	21.544	\pm	0.009	$&	$	<10^{-15	}$	&$	19.37	\pm	2.27	$&	0.50	&	$	<10^{-15	}$	\\
V* V344 Pup		&	6.89	&	1.06	&	4778	&	1.45	&$	11.757	\pm	0.001	$&		-		&	-			&	-	&		-		\\
V* V354 Lib 	 $\bigstar$ 	&	11.34	&	1.23	&	4439	&	0.37	&$	5.623	\pm	0.0004	$&	$	<10^{-15	}$	&$	13.30	\pm	1.36	$&	1.19	&	$	<10^{-15	}$	\\
    \end{tabular}
    \end{adjustbox}
    \centering
\end{table*}

\begin{table*}
\centering
    \caption*{\textit{(continued)}}
    \begin{adjustbox}{width=\textwidth}
    \begin{tabular}{lccccccccc}
    \hline\noalign{\smallskip}
    ID & $V$ & $B-V$ & $T_\mathrm{eff}$ & $\log L$    & $P_\mathrm{rot}$  & 
    FAP ($P_\mathrm{rot}$) &  $P_\mathrm{cyc}$  & $A_\mathrm{cyc}$ & 
    FAP ($P_\mathrm{cyc}$) \\
       &     &       &  [K]             & [$L_{\odot}$] &      [d]                 &     &  [yr]             &  [\%]       &      \\
    \hline\noalign{\smallskip}
V* V356 Pup 	 $\maltese$  $\dagger$ 	&	8.21	&	1.21	&	4445	&	2.12	&$	37.306			$&		-		&$	4.20	\pm	0.08	$&	0.30	&	$	10^{-13	}$	\\
V* V4091 Sgr		&	8.45	&	1.16	&	4527	&	1.56	&$	16.932	\pm	0.011	$&		-		&	-			&	-	&		-		\\
V* V4138 Sgr 	 $\bigstar$ 	&	6.74	&	1.02	&	4713	&	1.18	&$	61.933	\pm	0.067	$&	$	<10^{-15	}$	&$	5.35	\pm	0.09	$&	0.92	&	$	<10^{-15	}$	\\
V* V4139 Sgr		&	8.45	&	1.17	&	4363	&	1.66	&$	44.828	\pm	0.018	$&		-		&	-			&	-	&		-		\\
V* V414 Hya 	 $\bigstar$ 	&	8.90	&	1.06	&	4586	&	1.39	&$	18.750	\pm	0.004	$&	$	<10^{-15	}$	&$	5.51	\pm	0.31	$&	0.41	&	$	10^{-12	}$	\\
V* V592 Pup		&	7.86	&	0.61	&	6023	&	0.45	&$	1.967	\pm	0.0001	$&		-		&	-			&	-	&		-		\\
V* V764 Cen 	  	&	8.91	&	1.37	&	4541	&	1.52	&$	22.519	\pm	0.019	$&	$	<10^{-15	}$	&$	12.43	\pm	0.53	$&	1.25	&	$	<10^{-15	}$	\\
V* V789 Mon 	  	&	9.39	&	0.96	&	4648	&	-0.50	&$	1.401	\pm	0.0001	$&	$	<10^{-15	}$	&$	13.47	\pm	1.65	$&	0.53	&	$	<10^{-15	}$	\\
V* V829 Cen 	  	&	7.84	&	0.93	&	4614	&	1.32	&$	11.707	\pm	0.003	$&	$	<10^{-15	}$	&$	10.30	\pm	0.30	$&	0.41	&	$	<10^{-15	}$	\\
V* V841 Cen 	 $\bigstar$ 	&	8.57	&	1.06	&	4588	&	0.60	&$	5.997	\pm	0.001	$&	$	<10^{-15	}$	&$	4.32	\pm	0.52	$&	0.23	&	$	10^{-4	}$	\\
V* V851 Cen 	  	&	7.80	&	0.94	&	4700	&	0.74	&$	12.225	\pm	0.007	$&	$	10^{-9	}$	&$	13.60	\pm	0.17	$&	1.33	&	$	<10^{-15	}$	\\
V* V858 Cen 	  	&	10.45	&	0.88	&	4939	&	-0.41	&$	1.037	\pm	0.00002	$&	$	<10^{-15	}$	&$	20.05	\pm	0.96	$&	0.91	&	$	<10^{-15	}$	\\
V* V965 Sco 	 $\bigstar$ 	&	8.57	&	0.95	&	4707	&	1.81	&$	30.544	\pm	0.025	$&	$	<10^{-15	}$	&$	4.51	\pm	0.04	$&	0.43	&	$	<10^{-15	}$	\\
V* V991 Sco 	 $\dagger$ 	&	9.69	&	0.68	&	5498	&	0.28	&$	3.300			$&		-		&$	5.97	\pm	0.17	$&	0.23	&	$	<10^{-15	}$	\\
V* VV Mon	 $\dagger$ 	&	9.40	&	0.81	&	4978	&	1.05	&$	6.051			$&		-		&	-			&	-	&		-		\\
V* VX Pyx 	  	&	6.37	&	0.95	&	5046	&	1.77	&	-			&		-		&$	9.43	\pm	0.48	$&	0.33	&	$	<10^{-15	}$	\\
V* YZ Men 	 $\bigstar$ 	&	7.77	&	1.08	&	4594	&	1.50	&$	19.216	\pm	0.006	$&	$	<10^{-15	}$	&$	16.99	\pm	2.38	$&	1.31	&	$	<10^{-15	}$	\\
    \hline 
    \end{tabular}
    \end{adjustbox}
    \centering
\end{table*}


\label{lastpage}
\end{document}